\documentclass[aps, twocolumn, superscriptaddress, showpacs, nofootinbib, longbibliography]{revtex4-1}

\usepackage[utf8]{inputenc}
\usepackage[T1]{fontenc}
\usepackage{ae,aecompl} 
\usepackage{graphicx}
\usepackage{amsmath}
\usepackage{color}
\usepackage{amssymb}
\usepackage{latexsym}
\usepackage{wasysym}
\usepackage{psfrag}
\usepackage{ifthen}
\usepackage[citecolor=blue,colorlinks=true]{hyperref}
\usepackage{longtable}
\usepackage{float}
\usepackage{lineno}

\newcommand{\referee}[1]{{\color{black}#1}}
\newcommand{\refereeB}[1]{{\color{black}#1}}

\newcommand{\Tdrift}{T_\mathrm{drift}}
\newcommand{\Dag}{$\dagger$}
\usepackage{todonotes}
	
\begin{document}

\title{Search for gravitational waves from twelve young supernova remnants with a hidden Markov model in Advanced LIGO's second observing run}

  \let\mymaketitle\maketitle
  \let\myauthor\author
  \let\myaffiliation\affiliation
  \author{Margaret Millhouse}\affiliation{OzGrav, School of Physics, University of Melbourne, Parkville, Victoria 3010, Australia}
  \author{Lucy Strang}\affiliation{OzGrav, School of Physics, University of Melbourne, Parkville, Victoria 3010, Australia}
  \author{Andrew Melatos}\affiliation{OzGrav, School of Physics, University of Melbourne, Parkville, Victoria 3010, Australia}
 
\begin{abstract}

Persistent gravitational waves from rapidly rotating neutron stars, such as those found in some young supernova remnants, may fall in the sensitivity band of the advanced Laser Interferometer Gravitational-wave Observatory (aLIGO).  Searches for these signals are computationally challenging, as the frequency and frequency derivative are unknown and evolve rapidly due to the youth of the source.   A hidden Markov model (HMM), combined with a maximum-likelihood matched filter, tracks rapid frequency evolution semi-coherently in a computationally efficient manner. We present the results of an HMM search targeting 12 young supernova remnants in data from Advanced LIGO's second observing run.  Six targets produce candidates that are above the search threshold and survive pre-defined data quality vetoes.  However, follow-up analyses of these candidates show that they are all consistent with instrumental noise artefacts.

\end{abstract}

\maketitle

\section{Introduction}
Young supernova remnants (SNRs) hosting rotating neutron stars are promising candidates for the detection of continuous gravitational waves (GWs) by the advanced Laser Interferometric Gravitational-wave Observatory (aLIGO) \citep{Riles2013,harry2010advanced,LIGO-Scientific-Collaboration2015}.
Detection of transient GW events from mergers of compact binaries has now become routine~\cite{Catalog}.  Persistent, periodic GW signals have not yet been detected, but they are an attractive target, because the GW strain is proportional to the stellar ellipticity, which is determined partly by the nuclear equation of state~\cite{Riles2013}.  Motivated by the opportunity to do fundamental nuclear physics experiments, several groups have conducted continuous wave searches covering the whole sky~\cite{EinsteinatHomeO1, AEIAllSky, LIGOAllSkyCWO2} and various specific targets, e.g. known pulsars~\cite{O1NarrowBand, O2NarrowBand}, the Galactic center~\cite{GalacticCenter,Piccinni2019}, and young SNRs~\cite{YoungSNRO1,AEISNRs,EAHCasA,OwenO2SNR}, which are the subject of this paper.

Young neutron stars are especially likely to be non-axisymmetric, as any ellipticity produced during the violent birth of the star has had less time to relax by Ohmic, viscous, or tectonic processes \citep{Knispel2008,Chugunov_2010,Wette_2010}. Mass quadrupole emission (e.g. thermoelastic \citep{Ushomirsky2000,Johnson-McDaniel2013} or magnetic \citep{Cutler2002,Mastrano2011,Lasky2013} mountains) is expected to occur at the neutron star's rotational frequency, $f_*$, or $2f_*$.  Current quadrupole emission, e.g. from a pinned superfluid~\cite{Jones2010,Melatos2015} or r-modes~\cite{Caride2019}, is expected to occur at $f_*$ or approximately $4/3 f_*$ respectively.

Traditional searches are hampered by the computational cost of trialling a huge number of matched-filters, when the spin frequency and its evolution are rapid and unknown.
\refereeB{The computing cost for these searches scales as $f_\mathrm{max}^{2.2}a^{-1.1}T_\mathrm{obs}^4$~\cite{YoungSNRS6}, where $a$ is the age of the neutron star, $f_\mathrm{max}$ is the highest frequency in the search band, and $T_{\rm{obs}}$ is the total length of the observation.}
%The number of required templates scales as $T_\mathrm{obs}^{n(n+1)}$, where $n$ is the highest derivative $f_*^{(n)}$ in the phase model.  
This makes searches on long stretches of data (e.g. $T_{\rm obs} \gtrsim 1\,{\rm yr}$) with unknown frequency evolution for young neutron stars computationally infeasible.
Neutron stars are also subject to timing noise~\cite{Cordes_2013}, which causes the signal to wander stochastically.

In this paper, we present the results of a hidden Markov model (HMM) search for continuous waves first introduced by Suvorova et. al in 2016~\cite{Suvorova2018}, using open data from advanced LIGO's second Observing Run~\cite{GWOSC,O1O2GWOSC}.
The HMM is both robust against spin wandering and computationally cheap.

The paper is organized as follows.
In Sec.~\ref{Sec:PreviousSNR} we give an overview of the methods used in previous searches for GWs from SNRs.  In Sec.~\ref{Sec:HMM} we introduce the HMM, and describe how the HMM formalism is used in the search for continuous GWs.  Section~\ref{Sec:SearchParams} explains the methodology for selecting the search parameters for each SNR.  In Sec.~\ref{Sec:targets} we go over the selection of SNR targets, and in Sec.~\ref{Sec:threshold} we introduce the methods for selecting a threshold for detection.  Sec.~\ref{Sec:results} presents the results of our search, included the requirements for vetoing a potential candidate. We conclude in Sec.~\ref{Sec:conclusion}.

\section{Methodological Overview}
\subsection{Previous SNR searches}\label{Sec:PreviousSNR}
Three searches for continuous GWs from SNRs were performed in data from initial LIGO \cite{CasAS5,YoungSNRS6,Sun1987a,EAHCasA}. More recently, three searches have been performed for GW emission from young SNRs in Advanced LIGO's first and second Observing runs (O1 and O2, respectively) \cite{YoungSNRO1,OwenO2SNR,AEISNRs}.  No detections were reported, and upper limits were set on the maximum GW strain emitted by each target.  Because O1 and O2 are more sensitive than Initial LIGO, \cite{YoungSNRO1,OwenO2SNR} improve significantly upon the upper limits set in Ref. \cite{YoungSNRS6}.  

\subsection{Matched filter}
Some of the previous searches \cite{YoungSNRS6,YoungSNRO1,OwenO2SNR,CasAS5} used a coherent matched-filter test that was based on the maximum likelihood $\mathcal{F}$-statistic \cite{Fstat}.  The  $\mathcal{F}$-statistic also plays an important role in the HMM search in this paper.

\referee{In the $\mathcal{F}$-statistic formulation, the detector data $d(t)$, is modeled as a GW signal, $h(t)$, plus stationary noise, $n(t)$, or explicitly
\begin{equation}
  d(t) = h(t) + n(t).
\end{equation}
The log-likelihoods of the signal ($\mathcal{H}_1$) and null ($\mathcal{H}_0$) hypotheses respectively are given by
\begin{eqnarray}
  p(d|\mathcal{H}_1) & = &-\frac{1}{2}\left<d-h|d-h\right> \\
  p(d|\mathcal{H}_0) & = &-\frac{1}{2}\left<d|d\right>,
\end{eqnarray}
where $\left<x|y\right>$ is the noise-weighted inner product, defined as
\begin{equation}
  \left<x|y\right> =  4\Re \int_0^\infty \frac{\tilde{a}(f)\tilde{b}^*(f)}{S_n(f)}\mathrm{d}f.
  \label{Eq:nwip}
\end{equation}
Here $S_n(f)$ is the one-sided noise power spectral density, and a tilde denotes the Fourier transform. The log-likelihood ratio of of the signal $h(t)$ given the data $d(t)$ can then be written as
\begin{equation}
  \log\Lambda = \left<d|h\right>-\frac{1}{2}\left<h|h\right>.
  \label{Eq:logLambda}
\end{equation}

For a persistent GW signal of constant amplitude, $h(t)$ can be written as
\begin{equation}
  h(t) = \sum_{\mu=1}^4 A^\mu h_\mu
  \label{Eq:Hoft}
\end{equation}
where $A^\mu$ are the amplitudes associated with $h_\mu$.  The $h_\mu$ are linearly independent, and are given by
\begin{eqnarray}
h_1 = a(t)\cos\Phi(t) \\
h_2 = b(t)\cos\Phi(t) \\
h_3 = a(t)\sin\Phi(t) \\
h_2 = b(t)\sin\Phi(t)
\end{eqnarray}
with $\Phi(t)$ giving the phase of the GW at the detector, accounting for the Doppler modulation of the signal due to the movement of the Earth. The functions $a(t)$ and $b(t)$ are the antenna response functions of the detector, and are written out explicitly in~\cite{Fstat}.  The log-likelihood ratio in Eq.~\ref{Eq:logLambda} can then be expressed as
\begin{equation}
  \log\Lambda = A^\mu d_\mu - \frac{1}{2}A^\mu A^\nu \mathcal{M}_{\mu\nu}
  \label{Eq:logLambdaCW}
\end{equation}
with $d_\mu\equiv\left<d|h_\mu\right>$ and $\mathcal{M}_{\mu\nu}\equiv\left<h_\mu|h_\nu\right>$.  

The $\mathcal{F}$-statistic is a maximum likelihood estimator, obtained by maximizing Eq.~\ref{Eq:logLambdaCW} with respect to $A^\mu$, and is given by
\begin{equation}
  \mathcal{F} = \frac{1}{2}d_\mu\mathcal{M}^{\mu\nu}d_\nu.
\end{equation}
The random variable $2\mathcal{F}$ is drawn from a noncentral chi-squared distribution with four degrees of freedom:
\begin{equation}
  \chi^2(2\mathcal{F}|4,\rho_0^2).
  \label{Eq:FstatDist}
\end{equation}
The non-centrality parameter $\rho_0$ is the optimal matched-filter signal-to-noise ratio.
}

\referee{To compute the $\mathcal{F}$-statistic, we use the \texttt{ComputeFStatistic\_v2} function that is part of the LIGO Analysis Library~\cite{lalsuite}.  The details of this implementation can be found in Ref.~\cite{Fstat2}. This implementation combines data from both detectors.  The noise spectral density $S_n(f)$ in Eq.~\ref{Eq:nwip} is estimated from the median of nearby frequency bins.}

The $\mathcal{F}$-statistic template models the continuous GW signal as a sinusoid with slow frequency evolution given by
\begin{equation}
\refereeB{f(t_\mathrm{SSB}) = f_* + \dot{f_*}(t_\mathrm{SSB}-t_0)+\frac{1}{2}\ddot{f_*}(t_\mathrm{SSB}-t_0)^2}, % changed f_0 to f_* to avoid any possible confusion with f_0 used the viterbi section below
\label{Eq:Fstat_template}
\end{equation}
where $t_0$ is the time at the start of the observing period, \refereeB{and $t_\mathrm{SSB}$ is the time at the solar system barycenter}.
%The $\mathcal{F}$-statistic accounts for amplitude modulation arising from the movement of the Earth.
Eq.~\ref{Eq:Fstat_template} does not account for stochastic spin wandering on time scales of days to weeks, known as  timing noise \citep{Hobbs_2010,Shannon_2010,Ashton_2015}, which represents a major challenge for traditional $\mathcal{F}$-statistic searches. 
\refereeB{Additionally, the young neutron stars in this search may secularly spin down so rapidly that the template bank includes a wide range of $f_*,\dot{f_*},\ddot{f_*}$ even in the absence of spin wandering}%Additionally, the young neutron stars in this search may spin down so rapidly, that a template bank proportional to $\dddot{f_*}$ must be kept in Eq.~\ref{Eq:Fstat_template}
, leading to an unmanageable number of templates.
Consequently, previous young SNR searches only use some of the available data.  For example, O1 spanned 130 days, but the searched data in Ref.~\cite{YoungSNRO1} only ranged from 3 to 44 days in the 15 targets~\cite{YoungSNRO1}.  The more recent $\mathcal{F}$-statistic search in O2 data spanned 12 to 55 days depending on the target, and searched a frequency band of \referee{15} to 150 Hz~\cite{OwenO2SNR}.

An alternative to a fully coherent matched-filter search is to break the data into smaller segments and perform a semi-coherent analysis.  A number of semi-coherent analyses have been used in LIGO and Virgo searches for continuous GWs \cite{CrossCorr2008, Krishnan2004, Goetz2011}. In this paper we perform a semi-coherent search that uses an HMM to track the GW frequency. The HMM employs recursion to prune efficiently the exponentially large bank of templates required to capture rapid secular spin down or stochastic spin wandering.
  
\subsection{HMM}\label{Sec:HMM}
An HMM relates a finite set of unobservable (``hidden'') discrete state variables to a finite set of observables. \referee{In this search, the hidden variable is the true GW frequency, $f_*$, and the observable variable is the $\mathcal{F}$-statistic described in the previous section.  We divide the full stretch of data of length $T_\mathrm{obs}$ into smaller segments of length $\Tdrift$, calculate the $\mathcal{F}$-statistic for each segment for a set of trial frequencies, $f_0$\footnote{Here $f_0$ refers to the search frequency, i.e. the frequency in the argument of the $\mathcal{F}$-statistic, and $f_*$ refers to the true frequency of the neutron star itself.}, and find the most likely evolution of the frequency, over the total observation time.}

The set of hidden variables constitutes a Markov chain.  A Markov chain describes a state $q(t)$ that wanders among a set of discrete states, $\{q_{0}, q_{1},...q_{N_Q}\}$, with state transitions happening at discrete time steps $\{t_0,t_1,...t_{N_t}\}$.  In this search, $q(t)=f_*(t)$, the true GW frequency. 
 %\referee{In this GW search, we divide the full stretch of data of length $T_\mathrm{obs}$ into smaller segments of length $\Tdrift$ which represent our discrete time steps.  The hidden state is the true GW frequency, $f_0(t)$}.
 A Markov chain is memoryless, so the state at time $t_{i}$ depends only on the state at the previous time step, $t_{i-1}$. The probability of a transition from one state to another is given by the transition probability 
\begin{equation}
A_{q_jq_i} = P\left(q_j|q_i\right),
\end{equation}
with $q(t_{n+1}) = q_j$ for some $j$, and $q(t_n)=q_i$ for some $i$.
%$q_j=q(t_{n+1})$ and $q_i=q(t_n)$.  
\referee{In this search, we assume that from time step $t_n$ to time step $t_{n+1}$, the frequency either stays in its current state ($q_{j} = q_{i}$), moves up one frequency bin ($q_{j} = q_{i+1}$), or moves down one frequency bin with equal probability ($q_{j} = q_{i-1}$), viz.
\begin{equation}
A_{q_iq_i} = A_{q_iq_{i+1}} = A_{q_iq_{i-1}} = \frac{1}{3}.
\label{Eq:TransitionProb}
\end{equation}  
All other probabilities are zero\footnote{Because young SNRs are expected to spin down rapidly~\cite{YoungSNRO1,Sun2018}, another choice would be $A_{q_iq_i}=A_{q_iq_{i-1}} = \frac{1}{2}$. To maximize flexibility and robustness, we choose to use Eq.~\ref{Eq:TransitionProb}. The extra computational burden is minimal, as confirmed in previous studies~\cite{Sun2018,Suvorova2018}.}.  Analyzing the data in segments eliminates the need to explicitly search over $\dot{f_0}$ and $\ddot{f_0}$. The data segmentation also allows for a more flexible model of frequency evolution to account for stochastic spin wandering~\cite{Cordes1980,Hobbs2010,Melatos1997,Bildsten1997} and magnetic dipole braking simultaneously, which is hard to achieve economically with a low-order Taylor expansion. 
}

%An HMM relates a finite set of unobservable (``hidden'') discrete state variables to a finite set of observables. 
The observable $o(t)$ occupies one of the discrete states $\{o_{0}, o_{1},...o_{N_O}\}$.  The observable state is related to the hidden state by an emission probability defined by
\begin{equation}
L_{o_iq_j} = P\left(o_i|q_j\right),
\end{equation}
with $o(t_n)=o_i$ for some $i$, and $q(t_n) = q_j$ for some $j$.  
\referee{The observable in this search is the $\mathcal{F}$-statistic.  We calculate $\mathcal{F}(f_0)$ for each segment of length $\Tdrift$ (the recipe for setting $\Tdrift$ is described in Section~\ref{Sec:SearchParams}), at a frequency resolution of $\Delta f_0 = 1/(2\Tdrift)$.  The emission probability is given by \cite{Suvorova2018}
\begin{eqnarray}
L_{o(t)q_i} & = & P\left[o(t)|f_{0_i}\leq f_0(t) \leq f_{0_i}+\Delta f_0\right] \\
& \propto & \exp\left[\mathcal{F}(f_0)\right],
%P\left(o(t)|f_{0_i}\leq f_0(t) \leq f_{0_i}+\Delta f_0\right)
\end{eqnarray}
where $f_{0_i}$ is the value of $f_0$ in the $i^\mathrm{th}$ frequency bin, and the proportionality to the exponential follows from Eq.~\ref{Eq:FstatDist}.
}

Over some observation period we can find the most likely hidden state sequence, $Q^*$, given the observable state sequence, $O$ by maximizing
\begin{equation}
\begin{aligned}
P\left(Q|O\right) = & L_{o(t_{N_t})q(t_{N_t})}A_{q(t_{N_t})q(t_{N_t-1})}\times...\\
 & \times L_{o(t_1)q(t_1)}A_{q(t_1)q(t_0)}\Pi_{q(t_0)},
\end{aligned}
\label{Eq:PQO}
\end{equation}
with respect to $Q$. In Eq.~\ref{Eq:PQO}, $\Pi_q(t_0)$ is the prior probability that the state started at $q_i$ at $t=t_0$. \referee{As we do not know $f_0(t_0)$, the prior is uniform:
\begin{equation}
\Pi_{q(t_0)} = \frac{1}{N_Q}.
\end{equation}}
The maximization can be done with the Viterbi algorithm~\cite{Viterbi1967}, which uses dynamic programming to sample the $N_Q^{N_T}$ sequences $Q$ efficiently.

%Having outlined the HMM formalism, we now discuss how it is implemented in GW searches in Sec.~\ref{Sec:HMMGWs}.

%%%%%%%%%%%%%%%%%%%
% NOT USING SCORE ANY MORE
%  THIS IS GARBAGE NOW
%%%%%%%%%%%%%%%%%%%
%To claim a detection, we need to establish a detection statistic.  Following previous HMM GW searches, we use the Viterbi score, $S$, which is defined to be
%\begin{equation}
%S=\frac{\ln \delta_{q^*}(t_{N_T})-\mu_{\ln \delta(t_{N_T})}}{\sigma_{\ln \delta t_{N_T}}},
%\label{Eq:ViterbiScore}
%\end{equation}
%where $\delta_{q_i}(T_{N_T})$ is the maximum probability of the path that ends in the state $q_i$, the maximum of that path is $\delta_{q^*}(t_{N_T})$ (i.e. the optimal path), and
%\begin{equation}
%\mu_{\ln \delta(t_{N_T})} = N^{-1}_Q\sum_{i=1}^{N_Q}\ln \delta_{q_i}(T_{N_T}),
%\end{equation}
%\begin{equation}
%\sigma_{\ln \delta t_{N_T}}= N^{-1}_Q\sum_{i=1}^{N_Q}\left(\ln \delta_{{q}_{i(t_{N_T})}}-\mu_{\ln \delta(t_{N_T})}\right).
%\end{equation}
%The Viterbi score is a measure of the number of standard deviations by which the likelihood of the path of interest exceeds the mean likelihood of all paths.  The method for selecting a threshold Viterbi score for detection is discussed in \todo{put appropriate section here}.

\section{Parameters}\label{Sec:SearchParams}
In this section we again outline the procedure for setting the parameters for an SNR search, namely the frequency range and $\Tdrift$.

\subsection{Frequency Range}
The SNRs we are targeting in this paper do not contain electromagnetically observed pulsars, so $f_0(t)$ is unknown.  We must therefore search over a broad range of frequencies.  To set the frequency range, we demand that the indirect, age-based, spin-down upper limit on the GW strain lies above the strain sensitivity of the search.
For a neutron star of age $a$ at a distance $D$ that is spinning down purely due to GW radiation, the characteristic strain $h_0$ satisfies $h_0 \leq h_0^{\mathrm{max}}$ with \cite{Wette2008}
\begin{equation}
h_0^\mathrm{max} = 1.26\times10^{-24}\left(\frac{3.3\ \mathrm{kpc}}{D}\right)\sqrt{\frac{300\ \mathrm{years}}{a}}.
\label{Eq:indirectUL}
\end{equation}
On the other hand, \refereeB{assuming Gaussian noise}, the 95\% confidence upper limit on strain sensitivity for an incoherent search is analytically predicted to be (see Appendix E of~\cite{Sun2018})
\begin{equation}
\label{Eq:sensitivity}
h_0^{95\%}=\Theta S_n(f)^{1/2}\left(T_\mathrm{obs}\Tdrift\right)^{-1/4},
\end{equation}
where $\Theta\simeq 35$ is an empirical statistical factor \cite{Wette2008,YoungSNRS6}, and $S_n(f)$ is the one-sided noise spectral density.  
In this paper we search over all $f_0$ satisfying $h_0^\mathrm{max} >  h_0^{95\%}$ from Eqs.~(\ref{Eq:indirectUL}) and ~(\ref{Eq:sensitivity}).
%The frequency range depends on $\Tdrift$ through Eq.~(\ref{Eq:sensitivity}).

\subsection{$\Tdrift$}\label{Sec:Tdrift}
The segment length, $\Tdrift$, is selected to minimize the mismatch in the $\mathcal{F}$-statistic.  The mismatch is the fractional loss of signal power caused by the discretization of the parameters in the template set~\cite{Owen1996,Brady1998,Prix2007}.
Previous HMM searches for low-mass X-ray binaries set $\Tdrift$=10 days, the fiducial autocorrelation time scale for stochastic spin wandering in accreting systems~\cite{ScoX1O1,ScoX1O2,Middleton2020}.  An HMM has also been used to search for GWs from a long-lived remnant of a binary neutron star merger~\cite{GW170817PMLong},  which used a much shorter $\Tdrift$=1 second, as the remnant is possibly spinning down very rapidly.
In young SNRs hosting a non-accreting neutron star, stochastic spin wandering with an autocorrelation time-scale of days to weeks, known as timing noise in radio pulsar astronomy~\cite{Arzoumanian1994,Cordes1980}, must be weighed against rapid secular spin down.
 
As shown in detail in \cite{Sun2018}, for a neutron star with a spin-down rate of $\dot{f}_*$, in order to keep the $\mathcal{F}$-statistic mismatch below 0.2 \refereeB{when only searching over a constant $f_0$ (i.e. $\dot{f}_0=0$) in each coherent time segment,} we require $\Tdrift$ to satisfy
\begin{equation}
\Tdrift \leq \left(2|\dot{f}_*|\right)^{-1/2}.
\label{Eq:Tdriftmax}
%\Tdrift = \sqrt{\frac{1}{2|\dot{f_0}|}}.
\end{equation}
Because the targets in this paper do not have visible pulsars, the spin-down rate $\dot{f}_*$ is not known \emph{a priori}.  The range of $\dot{f}_*$ to be used in this search can be found by considering the possible ranges of the braking index, $n=f_0\ddot{f}_*/\dot{f}_*^2$.
For a neutron star of characteristic age $a=f_*/[(n-1)\dot{f}_*]$, we have 
\begin{equation}
-\frac{f_*}{(n_\mathrm{min}-1) a}\leq\dot{f_*}\leq-\frac{f_*}{(n_\mathrm{max}-1) a}
\end{equation}
where $n_\mathrm{min}$ and $n_\mathrm{max}$ are the minimum and maximum braking indices respectively.  Purely electromagnetic or gravitational braking implies $n=3$ and $n=5$ respectively.
Current observations imply $2\leq n \leq 7$ \cite{Melatos1997,Archibald2016}.  
In this work we assume $n=2$ conservatively to capture the widest possible range of signals, yielding from Eq.~\ref{Eq:Tdriftmax}:
\begin{equation}
\Tdrift = \left(\frac{a}{2f_*}\right)^{1/2}.
\label{Eq:Tdrift}
\end{equation}

We note that Eq.~\ref{Eq:Tdrift} depends on $f_*$, which we do not know \emph{a priori}.  One option is to vary $\Tdrift$ according to the search frequency, $f_0$, but this adds computational costs as well as additional trials factors.  In this work we use a single $\Tdrift$ per SNR target, which is the $\Tdrift$ that corresponds to the highest frequency where $h_0^\mathrm{max} > h_0^{95\%}$.

\subsection{Summary}
\begin{figure}[t!]
  \centering
  \scalebox{0.6}{\includegraphics{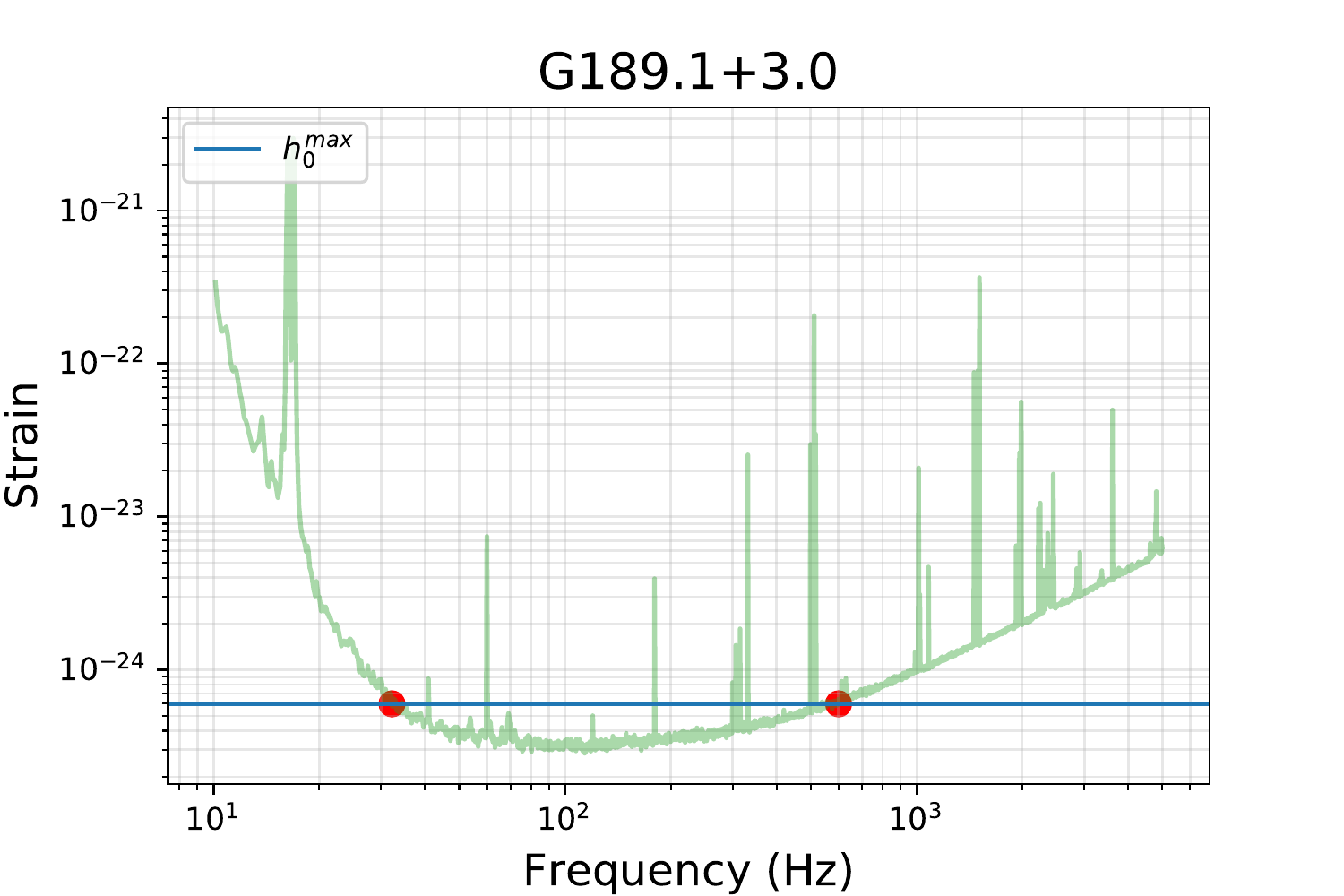}}
  \caption{Example of the predicted 95\% upper limit, $h_0^{95\%}$, from Eq.~\ref{Eq:sensitivity} (green), and the indirect upper limit, $h_0^\mathrm{max}$, for the SNR G189.1+3.0.  The red dots indicate $f_\mathrm{min}$ and $f_\mathrm{max}$.}	
  \label{Fig:sensitivityEx}
\end{figure}

The procedure for selecting $\Tdrift$ and the frequency bounds ($f_{\mathrm{min}}$, $f_{\mathrm{max}}$) for each SNR target is as follows:
\begin{itemize}
  \item Insert Eq.~\ref{Eq:Tdrift} into Eq.~\ref{Eq:sensitivity} to predict $h_0^{95\%}$ for $10\ \mathrm{Hz} < f_0 < 4000\ \mathrm{Hz}$, which is approximately the frequency band where LIGO is sensitive.
  \item Calculate the indirect upper limit $h_0^\mathrm{max}$ from Eq.~\ref{Eq:indirectUL}.
  \item Find the highest frequency obeying $h_0^\mathrm{max} > h_0^{95\%}$; call it $f_\mathrm{max}$.
  \item Using Eq.~\ref{Eq:Tdrift}, calculate $\Tdrift$ for $f_0=f_\mathrm{max}$.
  \item Insert $\Tdrift$ back into Eq.~\ref{Eq:sensitivity} and find the minimum frequency obeying $h_0^\mathrm{max} > h_0^{95\%}$;  call it $f_\mathrm{min}$.
\end{itemize}
Fig.~\ref{Fig:sensitivityEx} shows a predicted sensitivity curve, and indirect $h_0^\mathrm{max}$ for one example SNR.  The green curve shows Eq. \ref{Eq:sensitivity} for the calculated $\Tdrift$ of two hours.  The blue line is the indirect upper limit from Eq. \ref{Eq:indirectUL}, and the red points indicate $f_\mathrm{min}$ and $f_\mathrm{max}$.

\begin{table*}[ht]
\centering % used for centering table
\begin{tabular}{l c c c c c c c c c c c} % centered columns (4 columns)
\hline\hline 
SNR & Age & Distance & $f_\mathrm{min}$ & $f_\mathrm{max}$ & $\Tdrift$ & $h_0^\mathrm{max}$ & $h_0^{95\%}$ & RA & DEC & Sub-bands & Duty cycle \\ 
   & (kyr) & (kpc) & (Hz) & (Hz) & (hr) & $\times 10^{-25}$ & $\times 10^{-25}$ & (J2000) & (J2000) & & \\
\hline % inserts single horizontal line
G1.9+0.3~\cite{Reynolds2008,Green1984} & 0.1 & 8.5 & 35 & 122 & 1.0 (0.5) & 8.5 & 5.5 & 174846.9 & -271016 & 61 & 69\% \\ 

G18.9-1.1~\cite{Tullmann2010,Harrus2004} & 4.4 & 2 & 34 & 505 & 3.3 & 5.4 & 3.5 & 182913.1 & -125113 & 330 & 77\% \\ 

G65.7+1.2~\cite{Kothes2004,Kothes2008} & 20 & 1.5 & 42 & 335 & 8.5 & 3.4 & 2.7 & 195217.0 & 292553 & 205 & 83\% \\ 

G93.3+6.9~\cite{Jiang2007,Foster2003} & 5.0 & 1.7 & 32 & 600 & 3.1 & 5.9 & 3.5 & 205214.0 & 551722 & 397 & 77\% \\ 

G111.7-2.1~\cite{Tananbaum1999,Reed1995,Fesen2006} & 0.3 & 3.3 & 28 & 365 & 1.0 (0.6) & 12 & 5.2 & 232327.9 & 584842 & 236 & 69\%\\ 

G189.1+3.0~\cite{Swartz2015,Olbert2001} & 20 & 1.5 & 28 & 853 & 2.0 & 8.7 & 3.9 & 61705.3 & 222127 & 577 & 75\% \\ 

G266.2-1.2~\cite{Pavlov2001,Allen2015} & 5.1 & 0.9 & 18 & 840 &1.0 (0.4) & 14 & 5.8 & 85201.4 & -461753 & 575 & 69\% \\ 

G291.0-0.1~\cite{Slane2012,Moffett2001} & 1.2 & 3.5 & 36 & 471 & 1.7 & 5.9 & 4.0 & 111148.6 & -603926 & 305 & 73\% \\ 

G330.2+1.0~\cite{Park2006,Mcclure2001} & 1.0 & 5 & 46 & 288 & 2.1 & 4.5 & 3.9 & 160103.1 & -513354 & 169 & 74\% \\ 

G347.3-0.5~\cite{Mignani2008,Cassam1999,Wang1997} & 1.6 & 0.9 & 23 & 1747 & 1.1 & 20 & 4.6 & 171328.3 & -394953 & 1206 & 69\% \\ 

G350.1-0.3~\cite{Gaensler2008,Lovchinsky2011} & 0.6 & 4.5 & 36 & 474 & 1.2 & 6.5 & 4.4 & 172054.5 & -372652 & 307 & 70\% \\ 

G354.4+0.0~\cite{Roy2013} & 0.5 & 8 & 28 & 122 & 1.0 (0.4) & 14 & 6.0 & 173127.5 & -333412 & 66 & 69\% \\ 
\hline %inserts single line
\end{tabular}
\caption{SNRs targeted in this search.  For each target the table shows the astronomical parameters (RA, DEC, age, distance), search parameters ($f_\mathrm{min}$, $f_\mathrm{max}$, $\Tdrift$, and number of sub-bands), \refereeB{the indirect upper limit on the strain ($h_0^\mathrm{max}$) and predicted maximum sensitivity at $95\%$ confidence ($h_0^{95\%}$)}. \refereeB{For targets that are affected by the minimum $\Tdrift$ of 1 hour, we note in parentheses what the required $\Tdrift$ would be without the condition $\Tdrift \geq 1\ \mathrm{hr}$ imposed.}. The final column gives the duty cycle, or the percentage of $\Tdrift$ segments that had enough available data for at least the two SFTs required by the $\mathcal{F}$-statistic.} % title of Table
\label{Tab:searchParams} % is used to refer this table in the text
\end{table*}
\

\section{Implementation}
\subsection{Target selection}\label{Sec:targets}
In this work, we follow up on SNRs that have been targeted previously in LIGO data~\cite{YoungSNRS6,YoungSNRO1}.  Recently, Ref.~\cite{YoungSNRO1} searched O1 data for 15 young SNRs (as well as the neutron star Fomalhaut b).  These SNRs were selected from the Green catalog \citep{Green2019}. Another recent search has followed up on a subset of these targets~\cite{OwenO2SNR}. SNRs with central compact objects or pulsar wind nebulae are normally selected as they are likely hosts of neutron stars.

For each target, we select $\Tdrift$, $f_\mathrm{min}$, and $f_\mathrm{max}$ as described in Section~\ref{Sec:Tdrift}. The SNR targets and their respective search parameters are listed in Table~\ref{Tab:searchParams}. 
\refereeB{The $\mathcal{F}$-statistic ingests data in the form of short Fourier transforms (SFTs), and requires at least two SFTs~\cite{SFTs}.  This leads to the condition that $\Tdrift$ must be greater than twice the duration of the SFTs.  The typical SFT duration used in previous continuous GW searches is 30 minutes, which requires $\Tdrift \geq 1\ \mathrm{hour}$.  As a result, the predicted sensitivity for some targets from Ref. \cite{YoungSNRO1} cannot beat the indirect upper limit, i.e. those that are young and spinning down rapidly.  Additionally, $f_\mathrm{max}$ for some targets is bounded by the minimum $\Tdrift$ requirement rather than the sensitivity bounds in Sec. \ref{Sec:Tdrift}.  While it is possible in principle to produce SFTs of shorter durations, it requires extra computational time and data storage, and which exceed our computational resources.} 
%We impose the requirement $\Tdrift \geq 1\ \mathrm{hour}$, because the $\mathcal{F}$-statistic ingests data in the form of 30-minute short Fourier transforms (SFTs), and requires at least two SFTs~\cite{SFTs}.  
%\refereeB{For computational reasons, in this work we have not chosen to produce SFTs of shorter lengths, and leave that as a future exercise}.  

The parameter space of many targets span decades in Hz, so we split the search into sub-bands to facilitate data handling as in previous work \cite{ScoX1O2,ScoX1O1}.  In this work we search over sub-bands of 2 Hz.  This is wider than the sub-bands used previously (ranging from 0.606 Hz to 1.0 Hz) because rapid spin-down means the signal could transverse an entire sub-band during an interval of length $T_\mathrm{obs}$ if we use a width of 1 Hz or less. That is, there would be a high chance the signal would wander out of one sub-band, thereby decreasing the sensitivity of the search.
The sub-bands overlap, so that when a Viterbi path does straddle two sub-bands it is completely contained in one of the two.

\subsection{Detection statistic and threshold}\label{Sec:threshold}
Previous HMM searches used the Viterbi score~\cite{ScoX1O1,ScoX1O2} as the detection statistic.  The Viterbi score is the number of standard deviations that the log-likelihood of a path deviates from the average of all the other paths in a given sub-band, \refereeB{where the log-likelihood is the sum of the values of the $\mathcal{F}$-statistic at each step along the Viterbi path}.  The Viterbi score ceases to be useful when the number of frequency bins, $N_Q$, becomes comparable to the number of time steps, $N_T$.  
To understand why, consider how the Viterbi algorithm finds the optimal path.
By the principle of optimality \cite{Swart}, given an optimal path over $N_T$ time steps that ends in frequency bin $f_i$, the optimal path that ends in frequency bin $f_{i-1}$ (or $f_{i+1}$) is identical up to time step $N_T-1$. 
More generally, two paths terminating $j$ frequency bins apart have the same optimal subpath for time-steps $1<k<N_T-j$.
For $N_Q \gg N_T$, we have $N_T-j < 0$ for most paths, so most of the sub-optimal paths do not overlap.  For $N_Q\gtrsim N_T$ however, many of the final paths converge onto the same sub-optimal path.  If this path is a loud signal, it increases the mean of the log-likelihoods of all paths, thereby artificially decreasing the Viterbi score.
In short, in situations with  $N_Q\gtrsim N_T$, the Viterbi score for a true signal counterintuitively gets worse for longer observation times.  For this reason in this work we use the log-likelihood of the optimal path ending in each frequency bin as our detection statistic, unnormalized by the log-likelihoods of the neighboring paths.  We denote the log-likelihood as $\mathcal{L}$.

\begin{figure}[h]
  \centering
  \scalebox{0.6}{\includegraphics{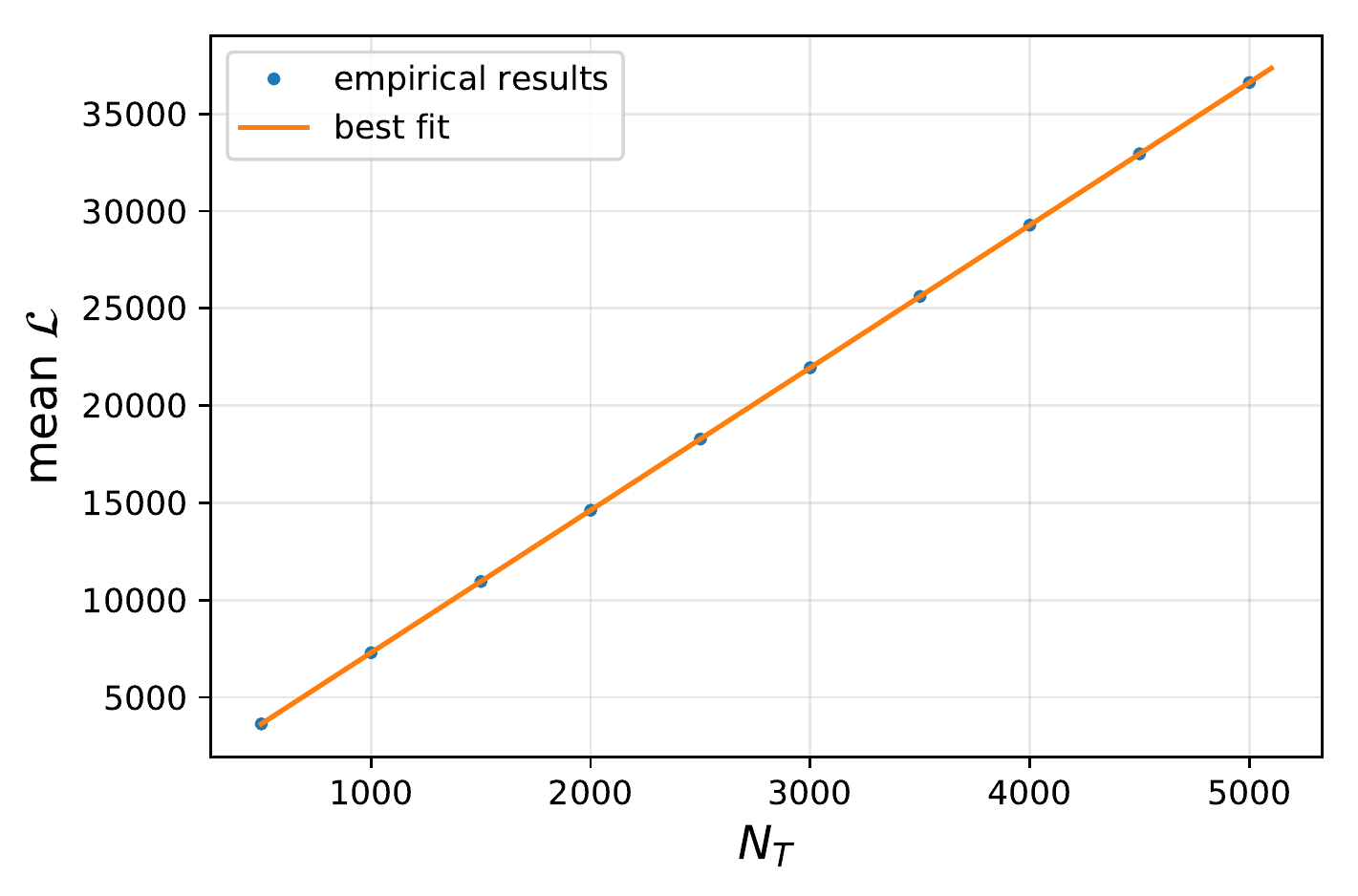}}
  \scalebox{0.6}{\includegraphics{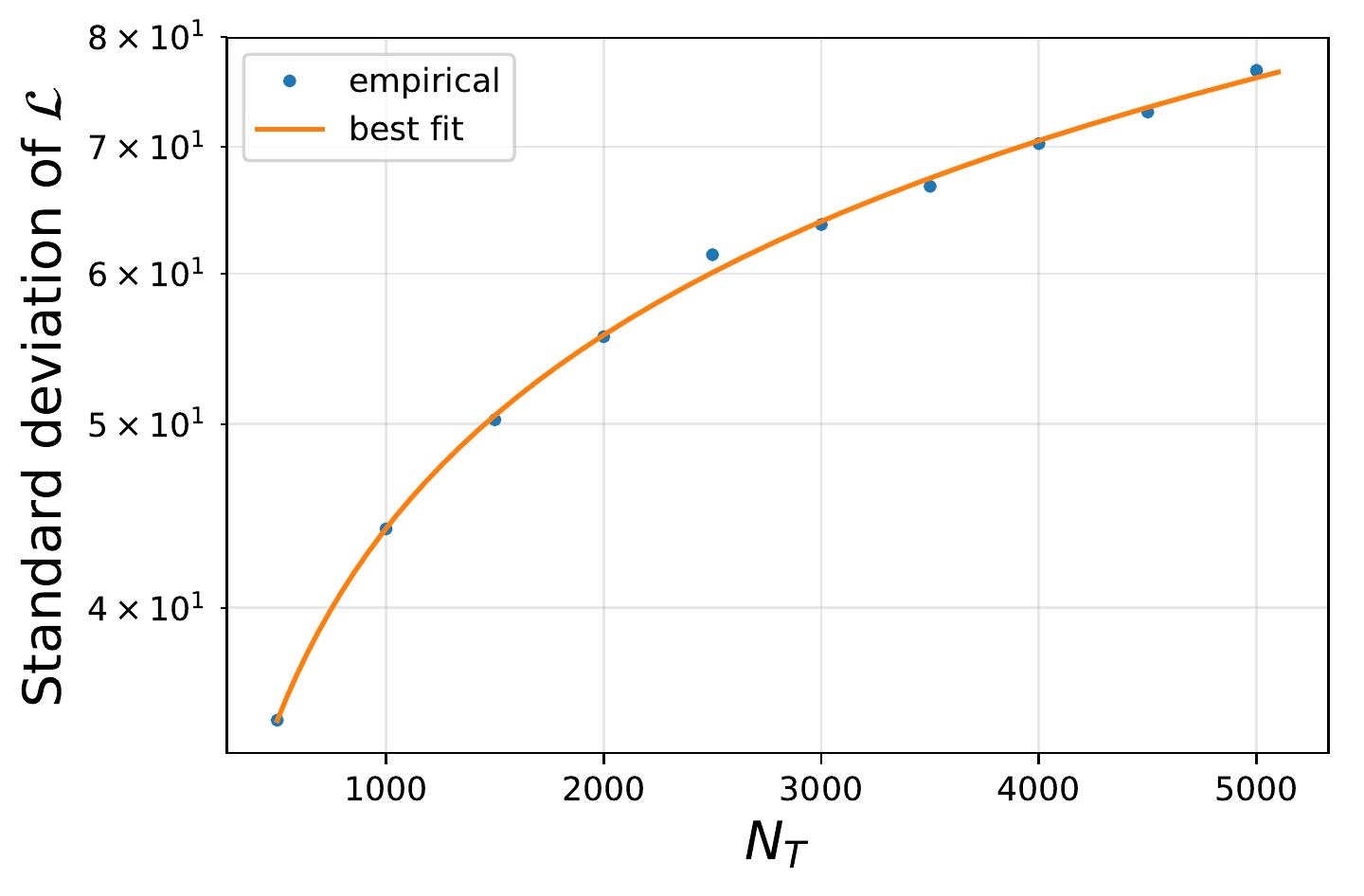}}
  \caption{The mean (top) and standard deviation (bottom) of $\mathcal{L}$ of the optimal path in Gaussian noise versus the number of time steps $N_T$.  The blue points are the empirical results. The orange curve is the best fit to those points.}
  \label{Fig:likelihood}
\end{figure}

The probability distribution function of $\mathcal{L}$ of the optimal path is not known analytically; see Section III C of~\cite{Suvorova2018} for details.  As verified empirically in Gaussian noise, the mean and standard deviation of $\mathcal{L}$ depend only on $N_T$ and scale in a well behaved manner.  Fig.~\ref{Fig:likelihood} shows the mean and standard deviation of the distribution of log-likelihoods in 100 realizations of Gaussian noise versus $N_T$ for $500\leq N_T\leq 5000$, relevant to the SNRs in this paper.  We find that the mean of $\mathcal{L}$ scales $\propto N_T$, and the standard deviation of $\mathcal{L}$ scales $\propto N_T^{0.34}$

\begin{table}[h]
\caption{Threshold and the number of outliers above that threshold before and after applying the data quality vetoes.} % title of Table
\centering % used for centering table
\begin{tabular}{l l c c} % centered columns (4 columns)
\hline\hline %inserts double horizontal lines
SNR  & $\mathcal{L}_\mathrm{th}$  & Outliers & Outliers \\ 
 & & (pre-veto) & (post-veto) \\
\hline % inserts single horizontal line
 G1.9+0.3  & 47752 & 32 & 0 \\
 G18.9-1.1 & 14830  & 100 & 2 \\
 G65.7+1.2  & 5761 & 45 & 4 \\
 G93.3+6.9 & 15156 & 125 & 1 \\
 G111.7-2.1  & 47771 & 51 & 0 \\
 G189.1+3.0  &  23227 & 115 & 3 \\
 G266.2-1.2 & 47783 & 124 & 3 \\
 G291.0-0.1 & 27243 & 65 & 0 \\
 G330.2+1.0 & 23346 & 32 & 0 \\
 G347.3-0.5  & 45290 & 227 & 5 \\
 G350.1-.03  & 47774 & 58 & 0 \\
 G354.4+0.0 & 47753 & 38 & 0 \\
\hline %inserts single line
\end{tabular}
\label{Tab:outliers} % is used to refer this table in the text
\end{table}

% Definition of Viterbi score in ScoX1O1 paper: number of standard deviations by which that path?s log likelihood exceeds the mean log likelihood of all paths

We use the scalings in Fig.~\ref{Fig:likelihood} to set the $\mathcal{L}$ threshold, $\mathcal{L}_\mathrm{th}$.  In this study we demand an overall false alarm probability of $\alpha_N=0.01$ for each target across all of the relevant sub-bands, the standard used in previous HMM searches~\cite{ScoX1O1,ScoX1O2}.  For each sub-band the desired false alarm probability $\alpha$ satisfies
\begin{equation}
\alpha_N = 1-(1-\alpha)^N
\label{Eq:totalFAP}
\end{equation}
where $N$ is the number of sub-bands multiplied by $N_Q$.

The thresholds obtained from the above procedure are shown in Table~\ref{Tab:outliers}.  The threshold range is $5761\leq \mathcal{L}_\mathrm{th}\leq47783$.   The threshold scales with the age of the SNR, so that targets of similar age have similar $\mathcal{L}_\mathrm{th}$, though targets with many sub-bands incur more trials, thus increasing $\mathcal{L}_\mathrm{th}$.

\subsection{Data}\label{Sec:data}
In this work, we search data from LIGO's second observing run, spanning 270 calendar days from November 2016 to August 2017.  A third detector, Virgo, joined O2 for the last month.  Due to the short duration of the Virgo run and its lower sensitivity, we analyze only data from the two LIGO detectors, Hanford and Livingston in this paper.  The strain data for O2 is publicly available from the Gravitational-wave Open Science Center~\cite{GWOSC,O1O2GWOSC,DataDOI}.

During O2 the detectors had periods of down-time.  There were two commissioning breaks during the run: an approximately two week period between December and January, and a break in May lasting 19 days for Livingston, and 31 days for Hanford.  In addition to these longer breaks, there were shorter periods of down time due to maintenance or environmental factors that brought the detectors out of lock.  As described in the previous section, the SFT data products require at least 30 minutes of data, so stretches of data shorter than this are not used in the analysis.  Furthermore, times in which the detector is known to not be properly operating in its nominal state are removed from the analysis~\cite{GW150914DQ,O1DQ}.  Because the $\Tdrift$ length periods used in this search are relatively short, there are sometimes $\Tdrift$ length periods where there is no analyzable data.  When this occurs, we fill in this period with a constant log-likelihood, as done in previous HMM searches~\cite{ScoX1O2}.  Accounting for missing SFTs, the effective duty cycles for each SNR are listed in Table~\ref{Tab:searchParams}.
%GW150914 paper DQ description: Data quality flags used in category 1 collectively indicate times when data should not be analyzed due to a critical issue with a key detector component not operating in its nominal configuration

\section{Results}\label{Sec:results}
\begin{figure}[t]
  \centering
  \scalebox{0.5}{\includegraphics{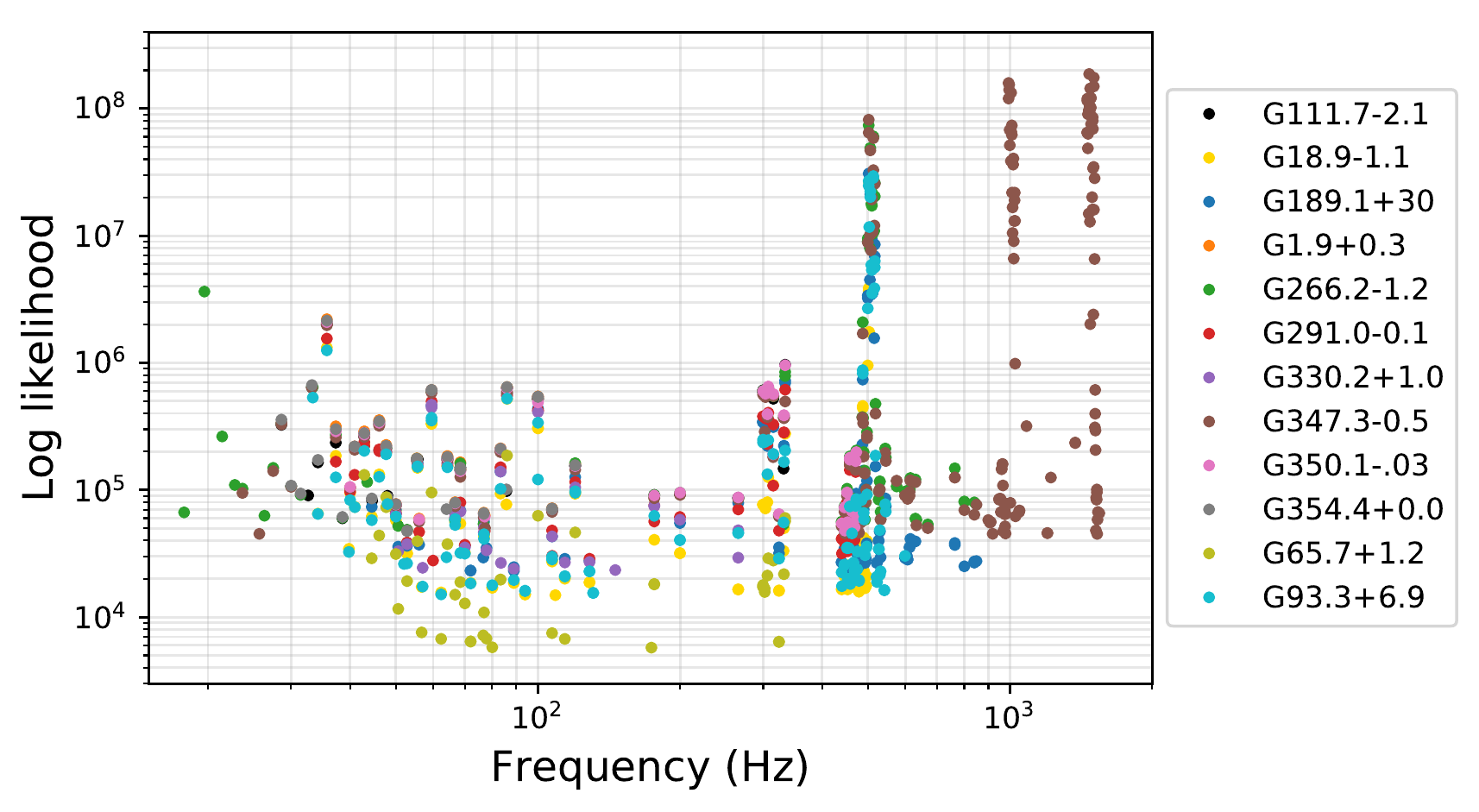}}
  \scalebox{0.5}{\includegraphics{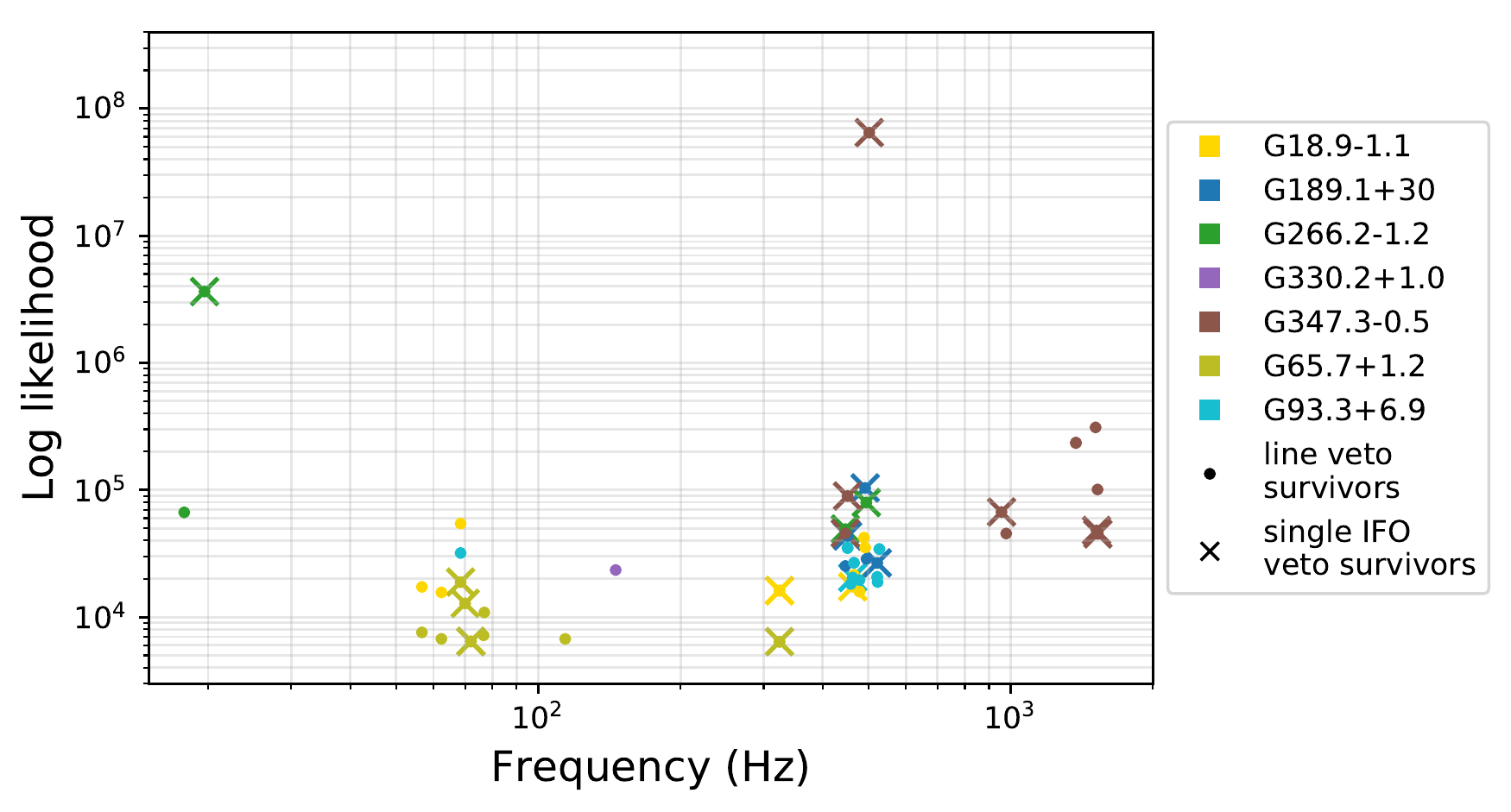}}
  \caption{Candidates whose log-likelihood exceeds the Gaussian threshold in Sec.~\ref{Sec:threshold}.  $\mathcal{L}$ is plotted against the terminating frequency of the associated Viterbi path, with points color-coded by their corresponding target (see legend at right). \emph{Top:} Candidates before vetoes. \emph{Bottom:} Survivors after the known line veto (circles), and remaining candidates after the single IFO veto (crosses). }
  \label{Fig:Outliers}
\end{figure}

All 12 of the targets in Table~\ref{Tab:searchParams} return Viterbi scores above the threshold defined in Sec.~\ref{Sec:threshold} in some sub-bands.  
The number of outliers per target is summarized in the third column of Table~\ref{Tab:outliers}.
$\mathcal{L}$ of every outlier is plotted versus frequency in Fig.~\ref{Fig:Outliers}, colored by target.

Several of the outliers are likely to occur because the detector noise is not Gaussian, as assumed when setting the threshold in Section~\ref{Sec:threshold}.  
To distinguish real signals from non-Gaussian noise, we pass the outliers through a set of vetoes used previously in published HMM searches \cite{ScoX1O1, ScoX1O2}.

\subsection{Vetoes}\label{Sec:vetoes}
Here we describe the vetoes in two categories. The motivating logic and implementation details for the vetoes are presented in Refs.~\cite{ScoX1O1,ScoX1O2}.
\begin{itemize}
\item \textbf{Instrumental noise lines}.  Narrowband instrumental noise artefacts known as ``lines'' are present in LIGO data at both interferometer sites \cite{LIGOLines}.  They are caused by suspensions vibrations, and the electrical power grid among other things.  We veto any candidate whose Viterbi path crosses the catalog of known instrumental lines \cite{GWOSC}.
\item \textbf{Single Interferometer Veto}.  An instrumental noise artefact that is present in one detector but not the other can artificially lift $\mathcal{L}$ from both detectors combined, $\mathcal{L}_{2\mathrm{ifo}}$,  above the threshold $\mathcal{L}_\mathrm{th}$.  To identify these false alarms, we rerun the search for each outlying sub-band in each interferometer separately.  If $\mathcal{L}$ in either interferometer (but not both) exceeds $\mathcal{L}_{2\mathrm{ifo}}$, we veto that candidate as an instrumental artefact.  If neither of the single-interferometer log-likelihoods exceeds $\mathcal{L}_{2\mathrm{ifo}}$, the candidate survives. 
\end{itemize}

Previous HMM searches have included a veto category in which the search is re-run, dividing the data into two segments.  A real signal should be significant in both segments and not turn on or off, although one can imagine exceptions, e.g. a transient r-mode~\cite{Caride2019}.  Previous searches however used the Viterbi score as a detection statistic~\cite{ScoX1O1,ScoX1O2}, which (when meeting the requirements described in Sec.~\ref{Sec:threshold}) is independent of $T_\mathrm{obs}$.  Since our detection statistic depends on $T_\mathrm{obs}$, we do not use this veto.

\begin{table*}[t]
% title of Table
\centering % used for centering table
%\begin{tabular}{c c c c c c l} 
\begin{tabular}{p{1.75cm} p{1.5cm} p{1.5cm} p{1.5cm} p{1.5cm} p{1.5cm} p{1.5cm} }
\hline\hline
SNR & $\mathcal{L}_{th}$ & Frequency & $\mathcal{L}$ & $\mathcal{L}$ & $\mathcal{L}$ & $\mathcal{L}$ \\ 
        &  & (Hz) &     & H1 only & L1 only & off-source  \\
\hline
G18.9-1.1 & 14830 & 323.994 & 16224 & 12342* & 8479 & 10340.6 \\ %14319 \\
- & - & 462.986 & 17321 & 14363* & 8467 & 17530\Dag \\ %16113\Dag
\hline
G65.7+1.2 & 5761 & 68.469 & 18848 & 6377 & 13890* & 8498\Dag \\ %12964\Dag \\
- & - & 69.997 & 12818 & 6412 & 5925 &  7275\Dag \\ %6474\Dag \\
- & - & 71.996 & 6440 & 3972 & 4337 &  4695 \\ %4907 \\
- & - & 323.977 & 6403 & 3898 & 3726 & 4484 \\ %6100\Dag \\
\hline
G93.3+6.9 & 15156 & 463.022 & 20483& 18235* & 9585 & 20683.6\Dag \\ %20489\Dag \\
\hline
G189.1+30 & 23227 & 451.503 & 43430 & 28129* & 12165 &  52394\Dag \\ %24844\Dag \\
- & - & 491.896 & 103623 & 65832* & 12212 & 98998\Dag \\ %61270\Dag \\
- & - & 521.749 & 26651 & 25177* & 13404 & 25308\Dag \\ %26959\Dag \\
\hline
G266.2-1.2 & 47783 & 19.650 & 3635140 & 372352 & 372352 & 1085260\Dag \\ %2516600\Dag \\
- & - & 446.677 & 49189 & 28319 & 22357 & 48633\Dag \\ %35833 \\
- & - & 494.676 & 79622 & 47087 & 47087 & 100052\Dag \\ %100219\Dag \\
\hline
G347.3-0.5 & 45290 & 446.703 & 45571 & 26376 & 21285 & 33606 \\ %33325 \\
- & - & 451.551 & 89539 & 59024* & 21161 & 52055\Dag \\ %51912\Dag \\
- & - & 501.859 & 64651000 & 37762400 & 3492760 & 26240600\Dag \\ %26757600\Dag \\
- & - & 956.293 & 67043 & 63642* & 21132 & 34872 \\ %62908\Dag \\
- & - & 1519.930 & 48015 & 43218* & 22481 & 44295 \\ %44627\Dag \\
\hline
\end{tabular}
\caption{Veto survivors.  The second through sixth columns list: the Gaussian threshold log-likelihood, the terminating frequency of the Viterbi path, the dual-interferometer $\mathcal{L}$, $\mathcal{L}$ from Hanford and Livingston only, and $\mathcal{L}$ of an off-source search.  An asterisk indicates that the event is much more significant in one interferometer than the other, and a dagger indicates that the off-source search also produces a candidate above the Gaussian threshold.  There are two survivors that are not marked with either a dagger or asterisk, one in G266.2-1.2 and one in G347.3-0.5.  The terminating frequencies of these candidates are similar (445.677 and 446.703), which suggests that these survivors are due to a common noise artefact.} 
\label{Tab:Outliers}
\end{table*}

\subsection{Survivors}
The fourth column of Table~\ref{Tab:Outliers} lists the veto survivors.  There are 18 spread across six SNRs. We report the terminating frequency of the Viterbi path, $\mathcal{L}$ of the original candidate, $\mathcal{L}$ of the single interferometer runs, and $\mathcal{L}$ of an off-source search.  

The off-source search is an additional follow-up procedure.  For all 18 outliers, we shift the right ascension by \refereeB{$10^{\prime}$ hours} while keeping all other search parameters fixed.  If the candidate is a true astrophysical signal, the resulting log likelihood should be consistent with Gaussian noise, with probability $1-\alpha$ of falling below $\mathcal{L}$ threshold.  If the off-source search exceeds $\mathcal{L}_\mathrm{th}$, there is likely to be an instrumental noise artefact in that band.  $\mathcal{L}$ for the single interferometer runs is included to show whether the candidate is much stronger in one detector than the other.   A candidate with a large asymmetry in the reported log-likelihoods from single interferometers can still be indicative of an instrumental noise artefact, even if neither log-likelihood exceeds $\mathcal{L}_{2\mathrm{ifo}}$ in the dual detector run as described in Section~\ref{Sec:vetoes}.
In particular, we note that $\mathcal{L}$ is mostly higher in the Hanford detector than the Livingston detector.  A real signal should not show this behavior, because in O2 Livingston was more sensitive than Hanford~\cite{Catalog}.

\begin{figure}
  \centering
  \scalebox{0.6}{\includegraphics{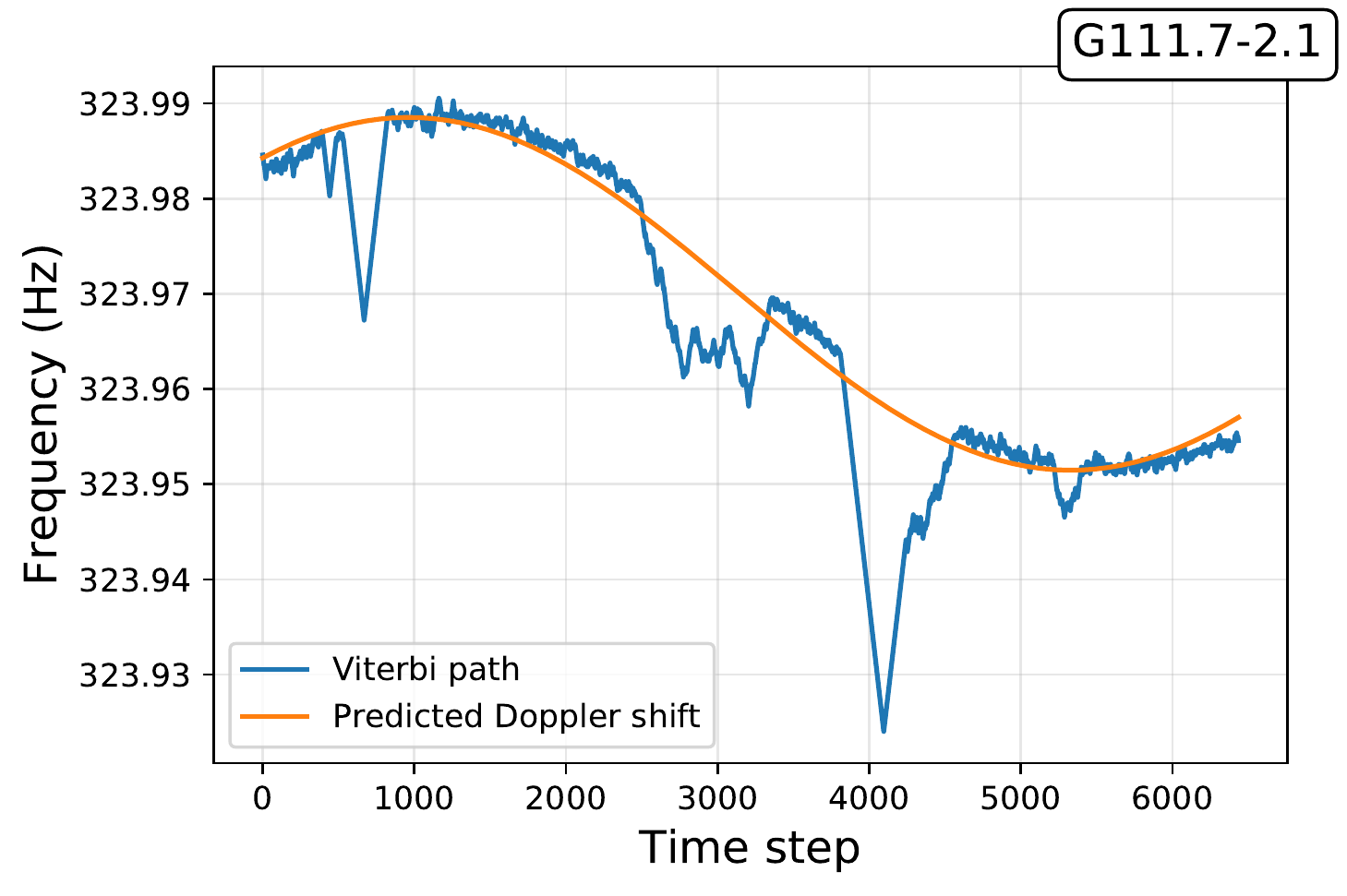}}
  \caption{HMM tracking of a Doppler-shifted instrumental line in the G111.7-2.1 search.  The orange curve shows the predicted Doppler shift of a stationary (in the detector frame) noise line processed by the $\mathcal{F}$-statistic. The blue curve shows the recovered Viterbi path. Note the magnified scale on the vertical axis.  }
  \label{Fig:Doppler}
\end{figure}

Several of the surviving outliers are close to known instrumental lines, even though outliers of similar frequency are vetoed via the known lines veto in one or more of the other targets. As the $\mathcal{F}$-statistic accounts for annual and diurnal Doppler modulation, lines that are stationary in the detector frame appear sinusoidal (with a period of a year) after passing through the $\mathcal{F}$-statistic.
Fig.~\ref{Fig:Doppler} shows the recovered Viterbi path for an outlier in SNR G111.7-2.1.  Overlaid on the Viterbi path is the predicted Doppler modulation of a stationary noise line as processed by the $\mathcal{F}$-statistic. The agreement is very good.

Next we briefly discuss all survivors.
\subsubsection{G18.9-1.1}
G18.9-1.1 has two candidates that survive the vetoes. Both show up more strongly in Hanford than Livingston.  

The candidate at 462.99 Hz has a log-likelihood of 12342 in H1, versus 8479 in L1.  This candidate also resurfaces as a significant outlier in the off-target search, indicating that it is not of astrophysical origin.

The candidate at 323.99 Hz is very close to an instrumental line, and similar candidates were vetoed for other targets.  Therefore we believe this outlier is caused by a noise artefact.

%The Viterbi paths for these two outliers are shown in Fig.~\ref{Fig:G18911Outliers}.

%\begin{figure}[h]
%  \centering
% \scalebox{0.45}{\includegraphics{G18911_subband_203}}
%  \scalebox{0.45}{\includegraphics{G18911_subband_300}}
%  \caption{HMM frequency tracks for the outliers in G18.9-1.1.}
%  \label{Fig:G18911Outliers}
%\end{figure}

\begin{figure}[h!]
  \centering
  %\scalebox{0.45}{\includegraphics{G65712_subband_18}}
  \scalebox{0.45}{\includegraphics{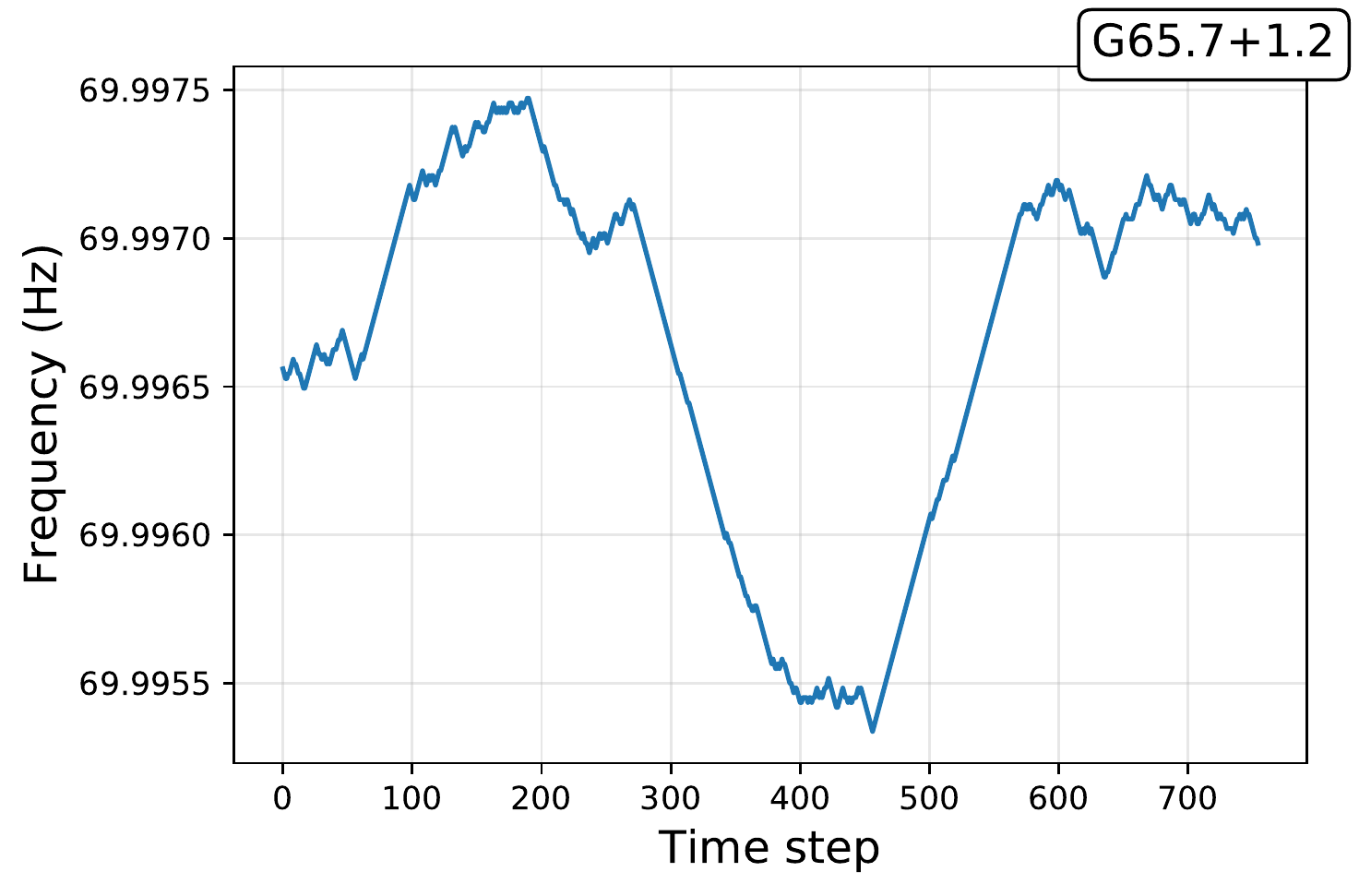}}
  \scalebox{0.45}{\includegraphics{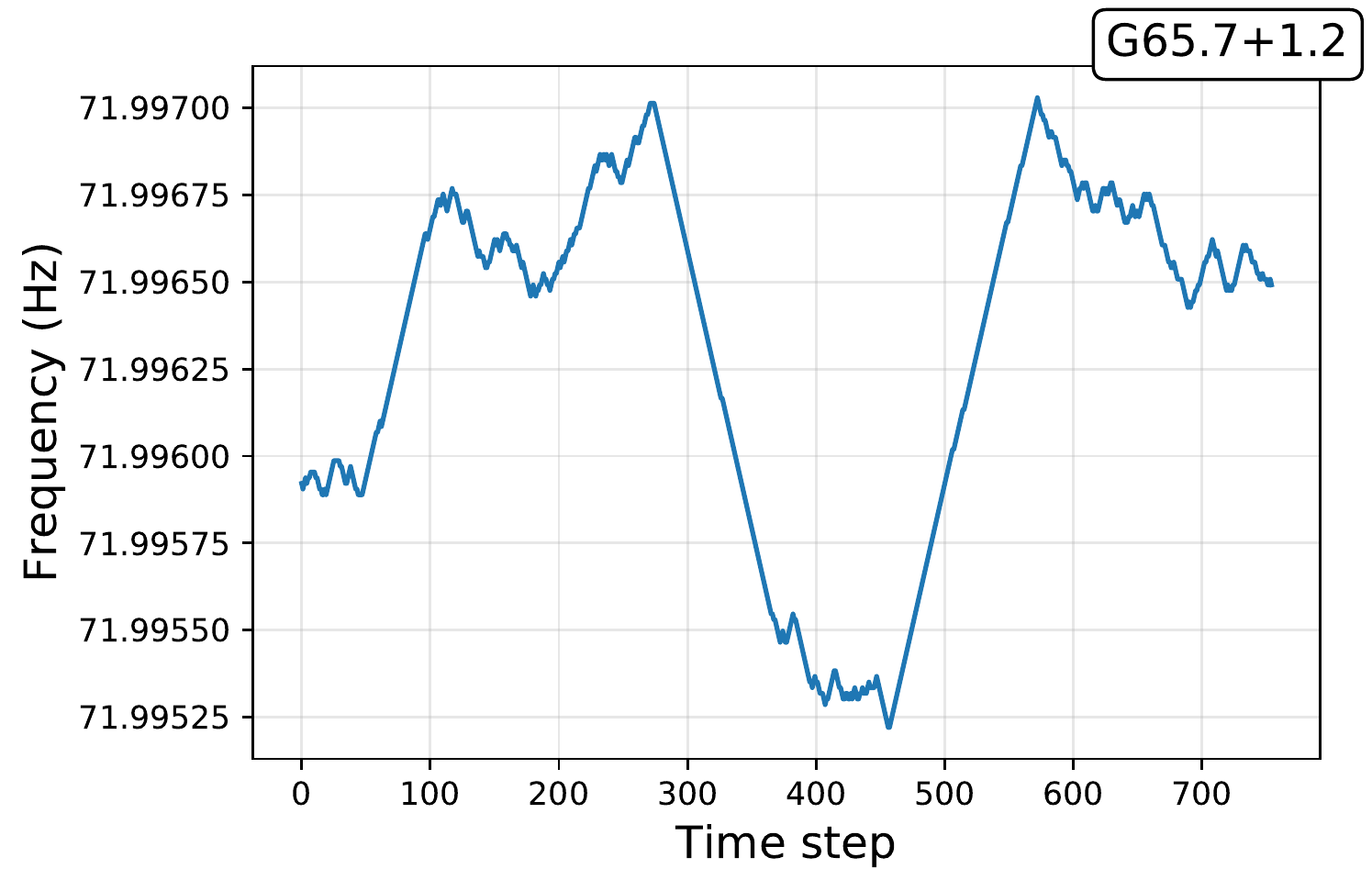}}
 % \scalebox{0.45}{\includegraphics{G65712_subband_197}}
  \caption{HMM frequency tracks for two of the the outliers in G65.7+1.2}
  \label{Fig:G65712Outliers}
\end{figure}

\begin{figure}[h!]
  \centering
  \scalebox{0.5}{\includegraphics{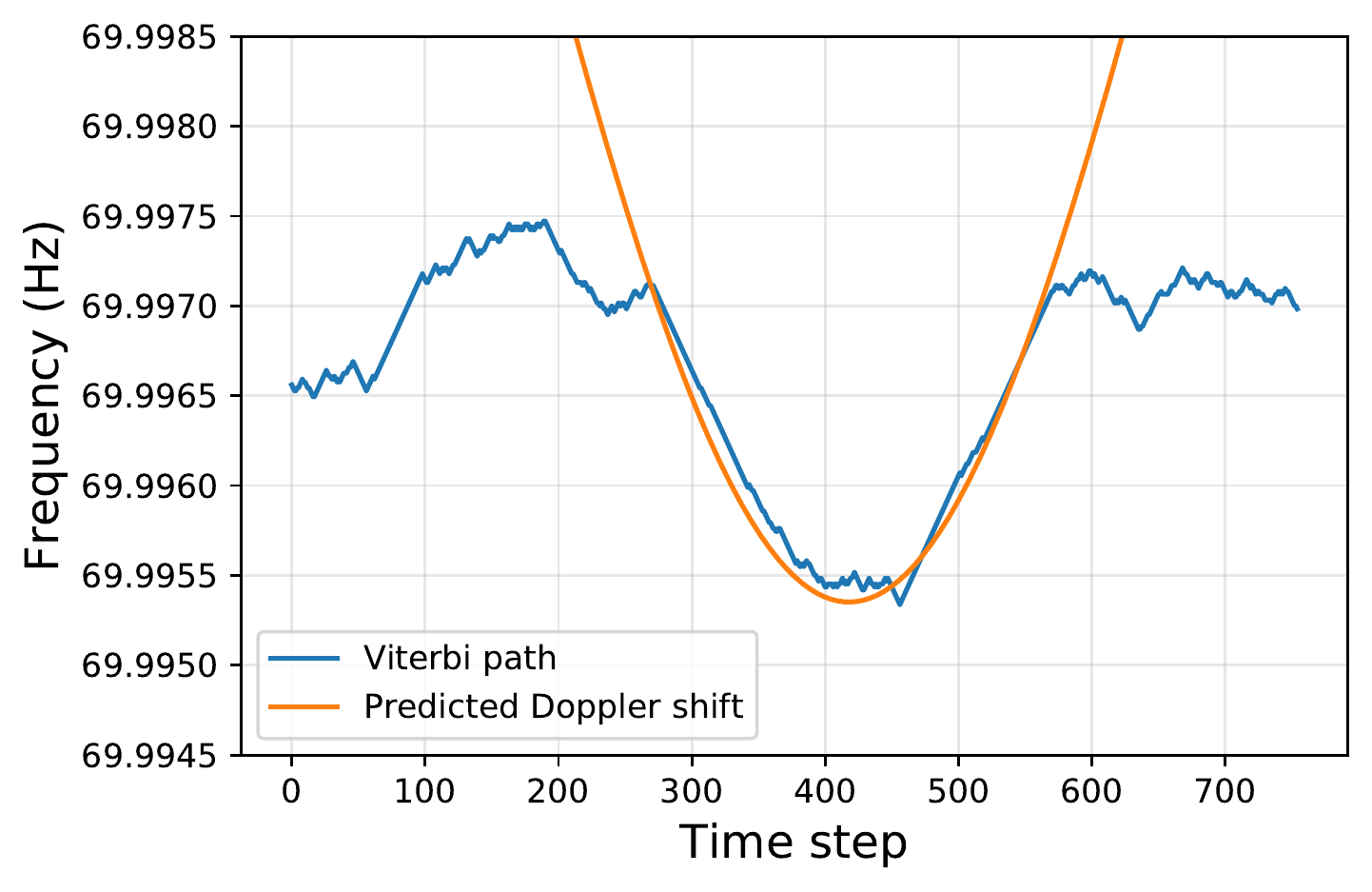}}
  \caption{The recovered Viterbi path for a candidate in G65.7+1.2 (blue), and the predicted Doppler modulation of a stationary noise line (orange).}
  \label{Fig:G65712Doppler}
\end{figure}

\subsubsection{G65.7+1.2}
There are four veto survivors in G65.7+1.2.  \refereeB{Two} of the candidates surpass $\mathcal{L}_\mathrm{th}$ in the off-source search, and one is much more significant in the Livingston detector than the Hanford detector. 

The candidate with a Viterbi path terminating at 71.996 Hz does not appear as an outlier in the off-source search, nor is it much more significant in one detector than in the other.  However, comparing the Viterbi path of this candidate to that of the candidate with a terminating frequency of 69.996 Hz, as shown Fig.~\ref{Fig:G65712Outliers}, we see that both paths exhibit similar behavior suggesting a common source \refereeB{e.g. a comb of noise lines~\cite{LIGOLines}}.  Overlaying the predicted Doppler modulation of a stationary noise line processed by the $\mathcal{F}$-statistic, we see a strong overlap with the Viterbi path as shown in Fig.~\ref{Fig:G65712Doppler}.  Hence we believe this survivor is from an instrumental noise artefact.

\refereeB{The remaining candidate with a terminating frequency of 323.977 Hz lies within $\approx 0.02$ Hz of eight other candidates vetoed in other targets. Hence it is likely that the candidate terminating at 323.977 Hz is a noise artefact.}  
% The targets with close candidates are: G111721, G18911, G266212, G291001, G347305, G350103, G93369, G189130
%Furthermore, we found via injection studies that for a stationary noise line in the detector if we repeat the search with $\mathcal{L}$}

%The Viterbi paths of all survivors are shown in Fig.~\ref{Fig:G65712Outliers}.

\subsubsection{G93.3+6.9}
G93.3+6.9 has one survivor, which is much more significant in Hanford than Livingston (18235 versus 9585), and very significant in the off-source search.  Hence, we do not believe it to be a real GW signal.

%The Viterbi path of the outlier is shown in Fig.~\ref{Fig:G93369Outliers}.

\subsubsection{G189.1+3.0}
There are three veto survivors in G189.1+3.0, with frequencies of approximately 451.50 Hz, 491.90 Hz, and 521.75 Hz.
All three are more significant in Hanford than in Livingston, and show up as significant candidates in the off-source search.  They are consistent with noise artefacts.

%The Viterbi paths for these candidates are shown in Fig.~\ref{Fig:G189130Outliers}.

\subsubsection{G266.2-1.2}
G266.2-1.2 has three survivors.  Two of these, at frequencies of 19.65 Hz and 494.68 Hz, are also significant in the off-source search. They are consistent with noise artefacts. 

\begin{figure}[h]
  \centering
  \scalebox{0.45}{\includegraphics{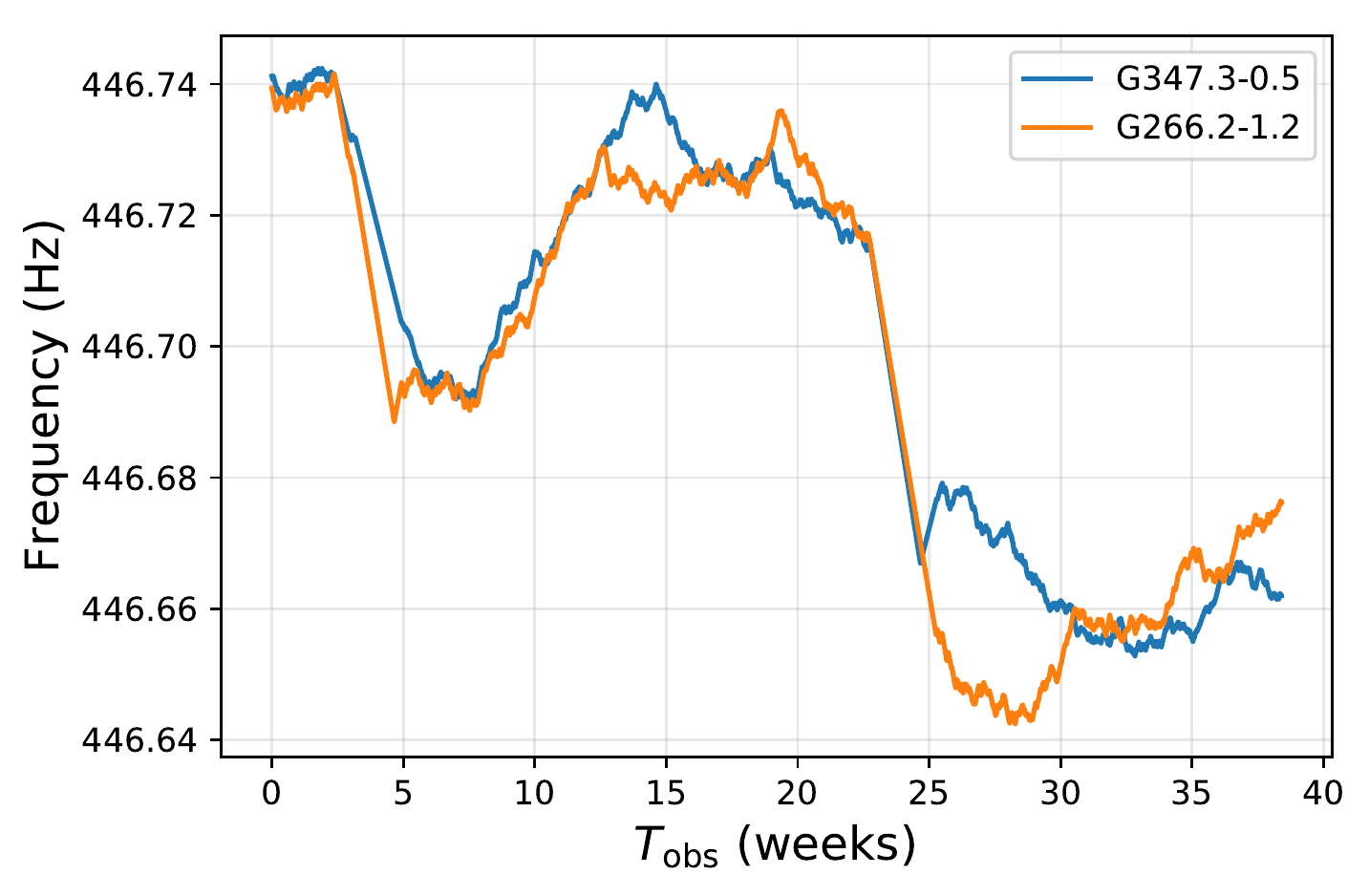}}
  \caption{HMM frequency tracks in the detector frame for two candidates of similar frequency in G266.2-1.2 and G347.3-0.5.  The two Viterbi paths are broadly consistent, indicating that these candidates arise from a common noise artefact.  Discrepancies in the paths arise from the different $\Tdrift$ and frequency band resolution used for the two targets.}
  \label{Fig:PathCompare}
\end{figure}

The third candidate is around 446.677 Hz.  The single interferometer and off-source log-likelihoods do not show anything that immediately indicates a noise artefact.  However, the target G347.3-0.5 independently generates a candidate at a very similar frequency (446.703 Hz). \refereeB{The HMM frequency paths of these candidates in the detector frame are shown in Fig.~\ref{Fig:PathCompare}; they are consistent with each other}. As there is no reason to believe two different SNRs emit GWs at the same frequency, the signal is unlikely to be astrophysical in origin.

%The Viterbi paths for these candidates are shown in Fig.~\ref{Fig:G266212Outliers}.

%\begin{figure}[h]
%  \centering
%  \scalebox{0.45}{\includegraphics{G93369_subband_301}}
%  \caption{HMM frequency tracks for the outlier in G93.3+6.9}
%  \label{Fig:G93369Outliers}
%\end{figure}

\subsubsection{G347.3-0.5}
G347.3-0.5 has five survivors.  Four of them show up more strongly in Hanford and/or have significant outliers in the off-source search.  

As mentioned above, the survivor at 446.703 Hz is very close in frequency to a survivor in the independent SNR G266.2-1.2.  Both are consistent with noise artefacts.

%The Viterbi paths for these candidates are shown in Fig.~\ref{Fig:G347305Outliers}.

\section{Conclusion}\label{Sec:conclusion}
In this work we present a search for continuous GWs from 12 young SNRs using an HMM combined with the maximum-likelihood $\mathcal{F}$-statistic.  This is one of the first searches for these targets in the LIGO O2 data set.  The semi-coherent nature of the HMM search confers computational savings, allowing us to use the entire stretch of O2 data. It also ensures that the search is robust to stochastic spin wandering on time-scales longer than $\Tdrift$, with $1\ \mathrm{hour} \leq \Tdrift \leq 8.5\ \mathrm{hours}$.

For each target, we select the search band and coherent analysis time, $\Tdrift$, to maximize the GW discovery potential.  After performing data quality vetoes, we find surviving candidates in six SNR targets.  Off-source searches and manual follow-up of these survivors indicates that all of them are due to instrumental noise artefacts, and not GWs.

\refereeB{Some previous HMM searches have placed upper limits on the strain of the GWs emitted by the target of the search~\cite{ScoX1O1,ScoX1O2}.  These limits follow from Monte Carlo simulations to determine the minimum detectable $h_0$ (at $95\%$ confidence).  Roughly 1000 signals of varying $h_0$ are injected into different noise realisations, and this process is repeated across a number of sub-bands. As this work involves 12 targets, each covering a wide frequency range with months of data, such an upper limit study becomes computationally expensive.  Additionally, the phase model in the HMM search is a random walk. Therefore any upper limits are not directly comparable with previous searches where the signal model is based on a Taylor expansion of the phase; in general, upper limits are conditional on the signal model in any search.  For these reasons, we do not produce upper limits in this work.}
%\refereeB{In this work, we do not place empirical upper limits on the GW strain for each SNR.  As this is a broad band search across 12 different objects, doing so would be a computationally intensive task, and so we leave this to future work.}

%\begin{figure}
%  \centering
%  \scalebox{0.45}{\includegraphics{G189130_subband_286}}
%  \scalebox{0.45}{\includegraphics{G189130_subband_314}}
%  \scalebox{0.45}{\includegraphics{G189130_subband_335}}
%  \caption{HMM frequency tracks for the outliers in G189.1+3.0.}
%  \label{Fig:G189130Outliers}
%\end{figure}

%\begin{figure}
%  \centering
%  \scalebox{0.45}{\includegraphics{G266212_subband_1}}
%  \scalebox{0.45}{\includegraphics{G266212_subband_300}}
%  \scalebox{0.45}{\includegraphics{G266212_subband_333}}
%  \caption{HMM frequency tracks for the outliers in G266.2-1.2.}
%  \label{Fig:G266212Outliers}
%\end{figure}

%\begin{figure}
%  \centering
%  \scalebox{0.45}{\includegraphics{G347305_subband_296}}
%   \scalebox{0.45}{\includegraphics{G347305_subband_299}}
%   \scalebox{0.45}{\includegraphics{G347305_subband_335}}     
%   \scalebox{0.45}{\includegraphics{G347305_subband_652}}
%   \scalebox{0.45}{\includegraphics{G347305_subband_1046}}
%  \caption{HMM frequency tracks for the outliers in G347.3-0.5.}
%  \label{Fig:G347305Outliers}
%\end{figure}

Just before submitting this manuscript, we became aware of a search for young SNRs by Lindblom and Owen~\cite{OwenO2SNR}. The two searches are similar in some ways, but there are four important differences:
\begin{enumerate}
  \item They are directed at overlapping but distinct sets of targets.  Specifically, targets searched in~\cite{OwenO2SNR} but not in this work are G15.9+0.2, G39.2-0.3, and G353.6-0.7.  Not included in~\cite{OwenO2SNR} are searches for the targets G111.7-2.1, G266.2-1.2, and G347.3-0.5 (though these targets were searched in~\cite{AEISNRs}).
  \item They search different bands.  The search presented in~\cite{OwenO2SNR} examines the band between 15 and 150 Hz for all targets in order to accommodate a fixed computational cost.  In this work the frequency band varies by target (see Table~\ref{Tab:searchParams}).  The narrowest frequency band searched is 35 to 122 Hz for G1.9+0.3, and the widest band is 23 to 1747 Hz for G347.3-0.5.  With two exceptions (G1.9+0.3 and G354.4+0.0), the bands in this search are wider.
  \item They analyze different volumes of data.  The search presented in~\cite{OwenO2SNR} uses a different observation time for each target.  The range of these observation times is 12 to 55.9 days.  The search presented in this paper uses all available O2 data, as outlined in Sec.~\ref{Sec:data}.
  \item The HMM search is semi-coherent and robust against spin wandering, whereas the work presented in~\cite{OwenO2SNR} uses a coherent matched-filter. 
\end{enumerate}
For all these reasons, the two analyses are complementary without being easily comparable. A comparative study of the sensitivities, even within their common band, is a tricky exercise to be attempted in future work.

LIGO is currently in its third observing run, O3, and and is expected to improve its sensitivity relative to O2.  More data at higher sensitivity increases our chances of making a detection of periodic GWs.  The HMM search can also be improved for rapidly spinning down SNR targets by tracking $\dot{f}_0$ as well as $f_0$~\cite{Sun2018}.

\section{Acknowledgements}
Parts of this research were conducted by the Australian Research Council Centre of Excellence for Gravitational Wave Discovery (OzGrav), through project number CE170100004.
This research has made use of data, software and/or web tools obtained from the Gravitational Wave Open Science Center (https://www.gw-openscience.org), a service of LIGO Laboratory, the LIGO Scientific Collaboration and the Virgo Collaboration. LIGO is funded by the U.S. National Science Foundation. Virgo is funded by the French Centre National de Recherche Scientifique (CNRS), the Italian Istituto Nazionale della Fisica Nucleare (INFN) and the Dutch Nikhef, with contributions by Polish and Hungarian institutes.

We thank the Continuous Wave Working Group of the LIGO Scientific Collaboration and Virgo Collaboration for their helpful discussion, and in particular Karl Wette and David Keitel for their helpful comments.  We also thank Hannah Middleton, Patrick Clearwater, Pat Meyers, and Lilli Sun for informative discussions.

\appendix*
\section{Hardware injections}
\begin{figure}
  \centering
  \scalebox{0.45}{\includegraphics{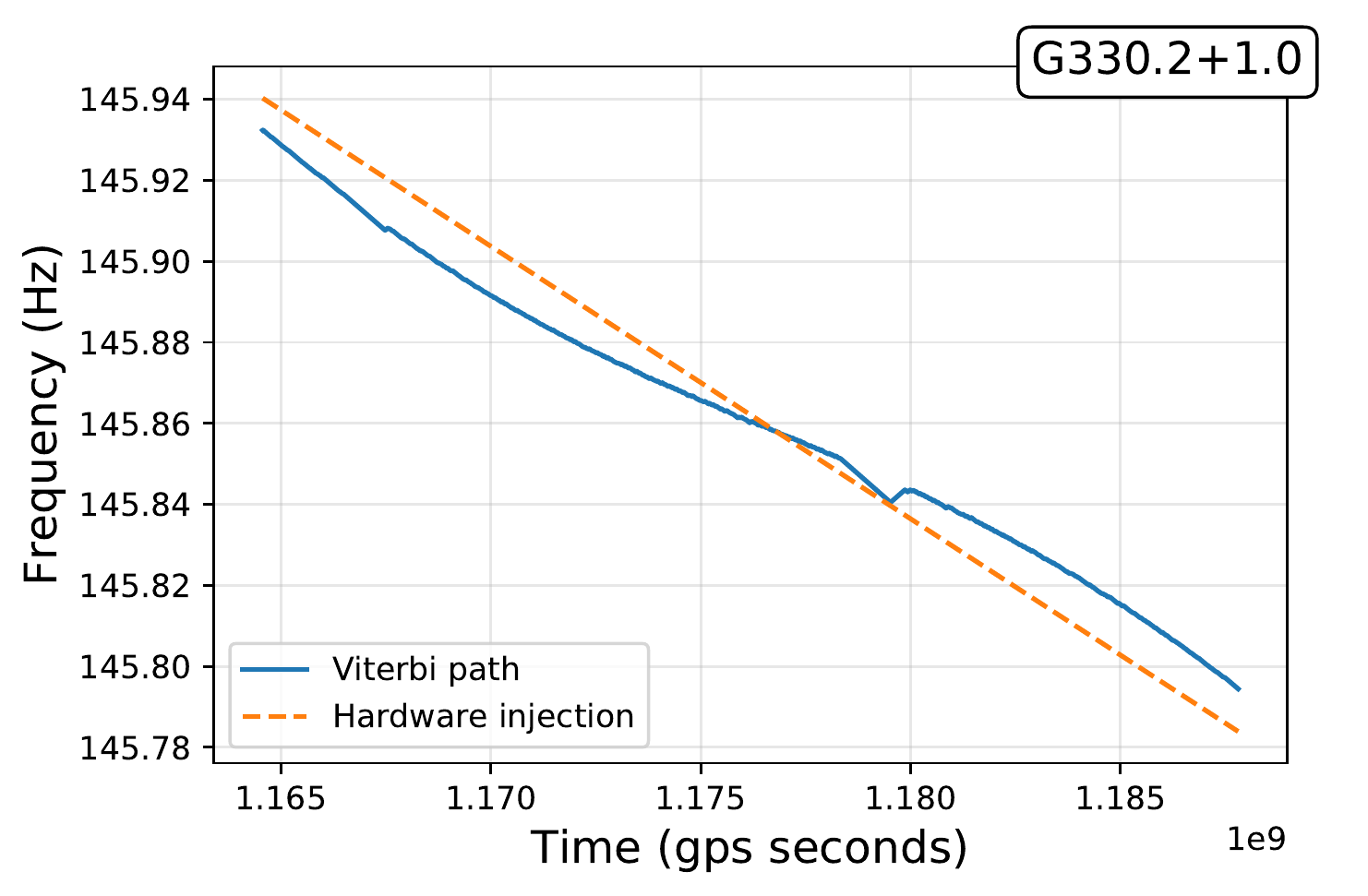}}
  \caption{Viterbi path corresponding to a hardware injection}
  \label{Fig:HW}
\end{figure}
To validate data analysis pipelines and calibration, simulated signals can be added to the LIGO detectors.  These are commonly called hardware injections.  In O2, injections were added to simulate GW signals from isolated rotating neutron stars~\cite{GWOSC,HWINJ}.  One such hardware injection is picked up by our search for the SNR target G330.2+1.0.  \refereeB{This candidate is from injected pulsar 6, as described in Ref.~\cite{LIGOAllSkyCWO2}.  Loud hardware injections have previously been detected at incorrect sky locations~\cite{YoungSNRO1}. As in this work, this particular injection was found in multiple targets in Ref.~\cite{OwenO2SNR}}.

The Viterbi path for this candidate, along with the frequency evolution of the hardware injection are shown in Fig.~\ref{Fig:HW}. $\mathcal{L}$ for the candidate, the single interferometer runs, and the off-source run are shown in Table~\ref{Tab:HW}.  We include the results to illustrate how a true GW signal would behave.
\begin{table*}
% title of Table
\centering % used for centering table
%\begin{tabular}{c c c c c c l} 
\begin{tabular}{p{1.75cm} p{1.5cm} p{1.5cm} p{1.5cm} p{1.5cm} p{1.5cm} p{1.5cm} }
\hline\hline
SNR & $L_{th}$ & Frequency & $\mathcal{L}$ & $\mathcal{L}$ & $\mathcal{L}$ & $\mathcal{L}$ \\ 
        &  & (Hz) &     & H1 only & L1 only & off-source  \\
\hline
G330.2+1.0 & 23346 &145.794 & 23452& 17350 & 16344 & 12112 \\
\hline
\end{tabular}
\caption{Details of hardware injection candidate.  The table lists the threshold log likelihood for each target with remaining outliers, the terminating frequency of the candidate path, $\mathcal{L}$ of the outlier, $\mathcal{L}$ of the H1 and L1 only runs, and $\mathcal{L}$ of an off source search.} 
\label{Tab:HW} % is used to refer this table in the text
\end{table*}

\bibliography{viterbiSNR}

\end{document}